# Efficient Controlled Bidirectional Quantum Secure Direct Communication using entanglement swapping in a network

Moein Sarvaghad-Moghaddam[1]

[1] Quantum Design Automation Lab, Amirkabir University of Technology, Tehran, Iran.

**Abstract:** In this paper, a novel controlled bidirectional quantum secure direct communication protocol (CBQSDC) is proposed. In this scheme, both users can transmit their secret messages directly and simultaneously under the permission of the controller. The proposed method use feature of entanglement swapping which is led to more efficiency (lowest number of quantum resources (EPR pairs)), security and the lowest number of security checking steps. Also, a generalization of the protocol is explained and a protocol of CBQSDC-based network is presented.

1. **Introduction**

Quantum cryptography [1] is one of the prominent issues in quantum information science which enables secure communication between legitimate participants using the physical laws of quantum mechanics. Applications of quantum cryptography includes Quantum Key Distribution (QKD) [2-5], Quantum Secret Sharing (QSS) [6-9], Quantum Secure Direct Communication (QSDC) [10-13], quantum coin tossing, multi-party computation (MPC) [14-16], randomness generation [17, 18] and so on.

One of the common protocols is QKD. In 1984, the first QKD protocol (BB84) [19] was proposed by Bennett and Brassard. QKD aims to establish a private key between two legitimate remote users. In this protocol, messages are not transmitted directly. Instead, secret keys are produced after the quantum transmission, and then they are used to encrypt the secret messages using one-time-pad. The encrypted messages are then sent over the classical channel.

In this paper, the focus is on QSDC protocols. In contrast to QKD, a quantum secure direct communication (QSDC) protocol directly transmits the secret messages without sharing a private key in advance. So, it is more useful into QKD, if the protocol has been unconditional security. Beige et al. [10] proposed the first QSDC protocol using single photon two qubit states. But eventually, the authors themselves realized the protocol is insecure. In [20] the well-known ping-pong protocol was presented which used EPR states for communication. It was proved later that this protocol was not secure in a noisy quantum channel [21]. After that, some QSDC protocols were proposed using different quantum states for communication [22-27]. However, they were not bidirectional, i.e., users could not exchange their secret messages simultaneously. For improving this drawback, the novel concept named as Quantum Dialogue (QD) was proposed by Nguyen [28] and Zhang et al. [29]. In QD [30], only one user can send his/her secret message to other at a time and vice versa. So, for transmitting the information in two directions, the protocol must be run twice. For instance, Fig. 1 shows a QD protocol using EPR pairs for communication between two legitimate users as introduced in [30].

In all of QSDC protocols, one of the problems is the security of the secret message transmitted in the quantum channel directly. So, without having security checking steps eavesdropper can successfully attack and steal some parts of the secret message. To solve this problem, Yan et al. [22] proposed a QSDC protocol based on EPR pairs and teleportation. In this scheme, without sending qubits and with local operations, secret messages are transmitted. So, this protocol, only need one step of security checking on the initial quantum channel. Therefore, if the secure quantum channel is used, this protocol is unconditionally secure (without security checking steps). Although, this protocol has some disadvantages. It needs an additional qubit for transmitting a message. Also, it is not bidirectional, i.e., it is one-way and only can transmit two-qubit from one side to other. In [31], Zhang and Man presented a bidirectional quantum communication protocol using entanglement swapping. As stated in the following, these protocols are not controllable. Another type of QSDC protocol is Controlled Quantum Secure Direct Communication (CQSDC). In this scheme, a sender can communicate with the receiver under the permission of a controller.

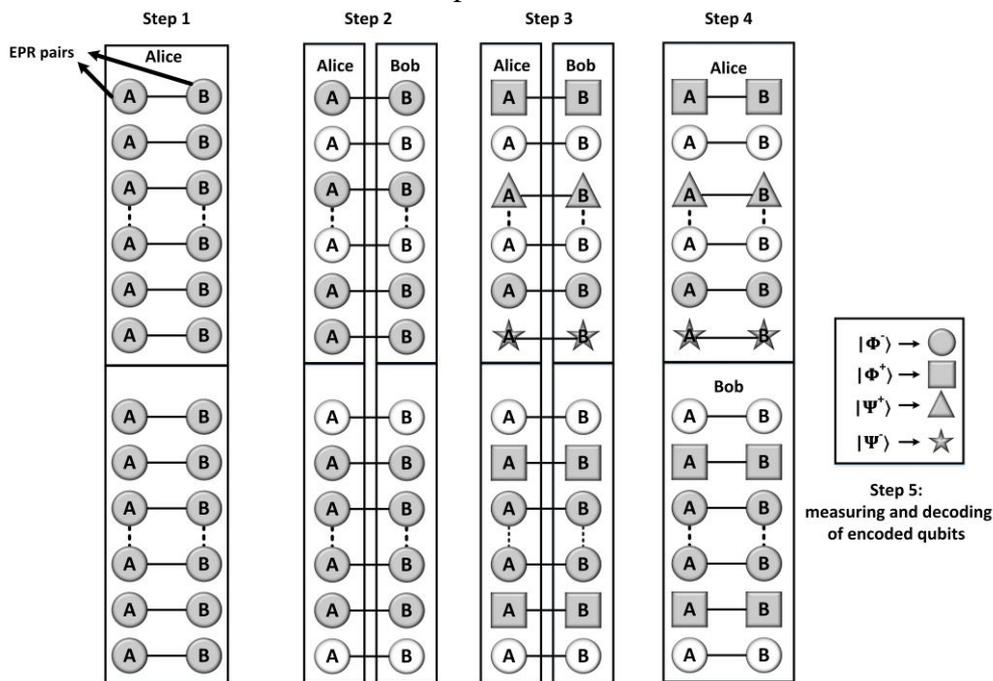

Fig. 1: Illustration of QSDC protocol with QD method as introduced in [30].

Generally, Controlled methods are safer than uncontrolled ones. As in uncontrolled ones, eavesdropper as a malicious user, Bob, comfortably can obtain the secret message with the scenario proposed in [32]. In [33] a CQSDC protocol is presented using GHZ state and feature of entanglement swapping. In this method, a six-qubit state is used for transmitting two message bits. This method is one-way and is not bidirectional. So, these methods used redundant resources and are not efficient. In [34], a CQSDC protocol is presented using GHZ-Like state and entanglement swapping. The method used in this protocol is the same as [33], and it only used GHZ-like state instead of GHZ state. Work [35] revisits one of previous works [36] and point out to the problem of leakage information in it. Then it presented a modified version of that work. This protocol uses W state. It is not bidirectional and controllable. This protocol needs to at least two-step of security; because secret messages are transmitted in the channel directly. Generally, above protocols are not optimal and use extra quantum resources.

In 2006 [37], a controlled bidirectional QDC (CBQDC) has been proposed by Man et al. in which users can transmit their secret messages with permission of controller at the same time. This protocol uses GHZ states for transmitting secret messages. It has low efficiency and uses extra quantum resources. Also, this protocol needs two steps of security checking for secure communication; unlike title in the paper, it is not bidirectional, and it is only a type of quantum dialogue (QD). Another disadvantage is that Charlie as controller needs to apply a three-qubit measurement, i.e. measurement based on GHZ states. Also, Ye et al. [38] proposed problem of information leakage in [37] which part of information can be detected in this protocol, and then they improve this problem and also proposed another protocol using Bell states. Unfortunately, both Ye et al.'s protocols still had the problem of the information leakage as stated in [39, 40]. In addition, other disadvantages of this paper are that the new method proposed in this paper use three security checking steps and unlike title in the paper, it is not bidirectional, and it is only type of quantum dialogue. In [32], the authors proposed an intercept-and-resend attack to Ye et al. protocol and then they proposed an improvement to solve intercept-and-resend attack and information leakage. This method has two disadvantage proposed in Ye et al. protocol, i.e. it uses three security checking steps and unlike title in the paper, it is not bidirectional, and it is only as QD.

In [41] a Controller-independent bidirectional quantum direct communication is proposed. In this protocol, two users have communication without the help of any third person so that any user cannot read the secret message without knowing the initial states generated by the other one. This work isn't bidirectional; because, it transmits the messages in two steps. Also, it at least needs two-step of security checking.

In this paper, a Controlled Bidirectional Quantum Secure Direct Communication (CBQSDC) is presented that each user can transmit two-bit message each to other in the same time using EPR states under the permission of the controller. In this protocol Controller only can monitor communication and cannot understand the messages. So it is a blind communication from the perspective of the controller. In this protocol, we improve the problems proposed in previous works. Because of using entanglement swapping, information is not directly transmitted between users; this is caused which the number of security checking steps is reduced. Then, a generalization of the protocol with GHZ state is presented. Finally, a protocol of CBQSDC-based network is presented. The proposed protocol effectively improves the quantum consumption resource into the previous ones.

The rest of this paper is organized as follows. In Section 2, the basic concept of entanglement swapping is presented. In Section 3 proposed CBQSDC protocol is presented. The security analyses are investigated in Section 4. In Section 5, the comparison and efficiency analysis are presented. Finally, Section 6 concludes the paper.

## 2. Entanglement Swapping

One of the most attractive features in quantum information is entanglement swapping [42-44]. Using this method, two particles that have never interacted can be entangled. For explaining this method, consider two EPR pairs with four particles labeled as 1 to 4. Qubits 1 and 2 are entangled together and also qubits 3 and 4 together. Also, qubits 1 and 4 are in the sender's possession and also qubits 2 and 3 in the receiver's possession. In these conditions, if one of the users performs a Bell state measurement on her/his particles. The whole state of system collapse as shown in Eq. (1). After that, particles 1 and 4 are entangled together and particles 2 and 3 together.

$$|\phi^+\rangle_{12} \otimes |\phi^+\rangle_{34} = \frac{1}{\sqrt{2}}(|00\rangle + |11\rangle)_{12} \otimes \frac{1}{\sqrt{2}}(|00\rangle + |11\rangle)_{34}$$
$$= \frac{1}{2}(|00\rangle_{12}|00\rangle_{34} + |00\rangle_{12}|11\rangle_{34} + |11\rangle_{12}|00\rangle_{34} + |11\rangle_{12}|11\rangle_{34})$$
$$= \frac{1}{2}(|00\rangle_{14}|00\rangle_{23} + |01\rangle_{14}|01\rangle_{23} + |10\rangle_{14}|10\rangle_{23} + |11\rangle_{14}|11\rangle_{23}) \quad (1)$$
$$= \frac{1}{2}(|\phi^+\rangle_{14}|\phi^+\rangle_{23} + |\phi^-\rangle_{14}|\phi^-\rangle_{23} + |\psi^+\rangle_{14}|\psi^+\rangle_{23} + |\psi^-\rangle_{14}|\psi^-\rangle_{23}).$$

As shown in Eq. (1), for instance, if the sender's result be $|\phi^+\rangle_{14}$, receiver's outcome will be $|\phi^+\rangle_{23}$ that are maximally entangled states. Also, similarity, entanglement swapping can be used in each different Bell state as the initial state.

Also, the entanglement swapping can be generalized for a network of users using GHZ states. So, there are 64 different states for entanglement swapping between two GHZ States. For instance, consider a network include of three users and a GHZ state. The above statements can be written for it in two way as the following:

1- Based on GHZ states:

$$|\phi_{000}\rangle \times |\phi_{000}\rangle = \frac{1}{2}((|000\rangle + |111\rangle) \times (|000\rangle + |111\rangle))_{abcdef}$$
$$= \frac{1}{2}(|000000\rangle + |000111\rangle + |111000\rangle + |111111\rangle)_{abcdef}$$
$$= \frac{1}{2}(|000000\rangle + |010101\rangle + |101010\rangle + |111111\rangle + \frac{1}{2}|000111\rangle$$
$$+ \frac{1}{2}|111000\rangle - \frac{1}{2}|000111\rangle - \frac{1}{2}|111000\rangle)_{adbecf}$$
$$= \frac{1}{2}(|\phi_{000}\rangle|\phi_{000}\rangle + |010101\rangle + |101010\rangle - \frac{1}{2}|000111\rangle - \frac{1}{2}|111000\rangle$$
$$+ \frac{1}{2}|010010\rangle + \frac{1}{2}|101101\rangle - \frac{1}{2}|010010\rangle - \frac{1}{2}|101101\rangle)$$
$$= \frac{1}{2}(|\phi_{000}\rangle|\phi_{000}\rangle + |\phi_{100}\rangle|\phi_{100}\rangle - \frac{1}{2}|000111\rangle - \frac{1}{2}|111000\rangle - \frac{1}{2}|010010\rangle$$
$$- \frac{1}{2}|101101\rangle)$$
$$= \frac{1}{2}(|\phi_{000}\rangle|\phi_{000}\rangle + |\phi_{100}\rangle|\phi_{100}\rangle + |\phi_{001}\rangle|\phi_{001}\rangle - |\phi_{101}\rangle|\phi_{101}\rangle)_{adbecf}$$

In the following table, eight different states of this entanglement swapping are seen.

| | |
|---|---|
| $\|\phi_{000}\rangle \times \|\phi_{000}\rangle = \frac{1}{2}((\|000\rangle + \|111\rangle) \times (\|000\rangle + \|111\rangle))_{abcdef}$ | $= \frac{1}{2}(\|\phi_{000}\rangle\|\phi_{000}\rangle + \|\phi_{100}\rangle\|\phi_{100}\rangle + \|\phi_{001}\rangle\|\phi_{001}\rangle - \|\phi_{101}\rangle\|\phi_{101}\rangle)_{adbecf}$ |
| $\|\phi_{000}\rangle \times \|\phi_{001}\rangle = \frac{1}{2}((\|000\rangle + \|111\rangle) \times (\|000\rangle - \|111\rangle))_{abcdef}$ | $= \frac{1}{2}(\|\phi_{000}\rangle\|\phi_{001}\rangle + \|\phi_{001}\rangle\|\phi_{000}\rangle + \|\phi_{100}\rangle\|\phi_{101}\rangle - \|\phi_{101}\rangle\|\phi_{100}\rangle)_{adbecf}$ |
| $\|\phi_{000}\rangle \times \|\phi_{010}\rangle = \frac{1}{2}((\|000\rangle + \|111\rangle) \times (\|100\rangle + \|011\rangle))_{abcdef}$ | $= \frac{1}{2}(\|\phi_{100}\rangle\|\phi_{000}\rangle + \|\phi_{101}\rangle\|\phi_{001}\rangle + \|\phi_{000}\rangle\|\phi_{100}\rangle - \|\phi_{001}\rangle\|\phi_{101}\rangle)_{adbecf}$ |
| $\|\phi_{000}\rangle \times \|\phi_{011}\rangle = \frac{1}{2}((\|000\rangle + \|111\rangle) \times (\|100\rangle - \|011\rangle))_{abcdef}$ | $= \frac{1}{2}(\|\phi_{101}\rangle\|\phi_{000}\rangle + \|\phi_{100}\rangle\|\phi_{001}\rangle + \|\phi_{000}\rangle\|\phi_{101}\rangle - \|\phi_{001}\rangle\|\phi_{100}\rangle)_{adbecf}$ |

$|\phi_{000}\rangle \times |\phi_{100}\rangle = \frac{1}{2}((|000\rangle + |111\rangle) \times (|010\rangle + |101\rangle))_{abcdef} = \frac{1}{2}(|\phi_{000}\rangle|\phi_{010}\rangle + |\phi_{001}\rangle|\phi_{011}\rangle + |\phi_{100}\rangle|\phi_{110}\rangle - |\phi_{101}\rangle|\phi_{111}\rangle)_{adbecf}$

$|\phi_{000}\rangle \times |\phi_{101}\rangle = \frac{1}{2}((|000\rangle + |111\rangle) \times (|010\rangle - |101\rangle))_{abcdef} = \frac{1}{2}(|\phi_{000}\rangle|\phi_{011}\rangle + |\phi_{001}\rangle|\phi_{010}\rangle + |\phi_{100}\rangle|\phi_{111}\rangle - |\phi_{101}\rangle|\phi_{110}\rangle)_{adbecf}$

$|\phi_{000}\rangle \times |\phi_{110}\rangle = \frac{1}{2}((|000\rangle + |111\rangle) \times (|110\rangle + |001\rangle))_{abcdef} = \frac{1}{2}(|\phi_{100}\rangle|\phi_{010}\rangle + |\phi_{101}\rangle|\phi_{011}\rangle + |\phi_{000}\rangle|\phi_{110}\rangle - |\phi_{001}\rangle|\phi_{111}\rangle)_{adbecf}$

$|\phi_{000}\rangle \times |\phi_{111}\rangle = \frac{1}{2}((|000\rangle + |111\rangle) \times (|110\rangle - |001\rangle))_{abcdef} = \frac{1}{2}(|\phi_{100}\rangle|\phi_{011}\rangle + |\phi_{101}\rangle|\phi_{010}\rangle + |\phi_{000}\rangle|\phi_{111}\rangle - |\phi_{001}\rangle|\phi_{110}\rangle)_{adbecf}$

2- Based on Bell states:

GHZ states can be changed to Bell states with entanglement swapping as shown in the following example:

$$|\phi_{000}\rangle \times |\phi_{000}\rangle = \frac{1}{2}((|000\rangle + |111\rangle) \times (|000\rangle + |111\rangle))_{abcdef}$$
$$= \frac{1}{2\sqrt{2}}(|\phi^+\rangle|\phi^+\rangle|\phi^+\rangle + |\phi^-\rangle|\phi^+\rangle|\phi^-\rangle + |\phi^+\rangle|\phi^-\rangle|\phi^-\rangle$$
$$+ |\phi^-\rangle|\phi^-\rangle|\phi^+\rangle + |\phi^+\rangle|\psi^+\rangle|\phi^+\rangle + |\phi^-\rangle|\psi^-\rangle|\phi^+\rangle - |\phi^+\rangle|\psi^-\rangle|\phi^-\rangle$$
$$- |\phi^-\rangle|\psi^+\rangle|\phi^-\rangle)_{adbecf}$$

### 3. Proposed method without network

In the proposed protocol, users can transmit their secret message to each other, simultaneously, under the control of Charlie – who acts as a controller – with one entangled state as a quantum channel. This scheme can be performed in five steps. In the first step, a quantum channel is created by Charlie. Checking security of quantum channel is done in the second step. In the third step, users encode messages, and secret dialogue is done. Other steps discuss decoding process and extraction of the message.

**Step 1**: **Distribution** The controller -Charlie- randomly prepares $2N + c$ Bell states from the set of four state $\{\phi^+, \phi^-, \psi^+, \psi^-\}_{A_x B_x}$ as shown in Eq. (2).

$$
\begin{aligned}
|\phi^+\rangle &= \frac{1}{\sqrt{2}}(|00\rangle_{AB} + |11\rangle_{AB}) = \frac{1}{\sqrt{2}}(|++\rangle_{AB} + |--\rangle_{AB}), \\
|\phi^-\rangle &= \frac{1}{\sqrt{2}}(|00\rangle_{AB} - |11\rangle_{AB}) = \frac{1}{\sqrt{2}}(|+-\rangle_{AB} + |-+\rangle_{AB}), \\
|\psi^+\rangle &= \frac{1}{\sqrt{2}}(|01\rangle_{AB} + |10\rangle_{AB}) = \frac{1}{\sqrt{2}}(|++\rangle_{AB} - |--\rangle_{AB}), \\
|\psi^-\rangle &= \frac{1}{\sqrt{2}}(|01\rangle_{AB} - |10\rangle_{AB}) = \frac{1}{\sqrt{2}}(|+-\rangle_{AB} - |-+\rangle_{AB}).
\end{aligned} \quad (2)
$$

Where $N$ is the number of qubits used for encoding the message, $c$ is the numbers of qubits used for the security checking and $x = 1,2,...,2N + c$. Then, Charlie distributes the qubits $A_x$ to Alice and $B_x$ to Bob and then both of them acknowledge the receipt of their respective qubits.

**Step 2**: **Channel verification** After Alice and Bob confirm that they received sequences, Alice executes security checking step as that she randomly selects $c$ 'checking' particles and measure them randomly in either *X* or *Z* basis. After that, she publishes the positions, measurement bases and measurement results to Bob by classical channel. Bob perform the same measurement in the same bases with Alice in corresponding particles. Alice also sends positions of this particles to Charlie. Then, Charlie notify the initial states of these 'checking' pairs to Alice and Bob. Afterwards, they compare their measurement results. If there is no eavesdropper, measurement results related to users must be correlated. If the error rate goes beyond the threshold, they will abort the communication, else they will continue to the next step.

**Step 3**: **Encoding**: After verifying the security of quantum channel, from the remaining $2N$ pairs, Alice and Bob randomly and according to a before conventional together, divide particles for transmitting their secret messages to each other. For example, Alice and Bob may simply select odd and even pairs, respectively. Then, they randomly create two-particle groups $((A_i, B_i), (A_j, B_j))$ in their sequences. For example, Alice (Bob) may group particles $\{P_1(A), P_2(A)\}$ ($\{(P_1(B), P_2(B)\}$) in Group 1, particles $\{P_3(A), P_4(A)\}$ ($\{P_3(B), P_4(B)\}$) in Group 2 and so on. Then, in order to encode their two-bit secret message on the i-th group, Alice and Bob performs the following operation on their corresponding group,

$$f(p,q,r,s) \to ((\sigma_x^p \sigma_z^q)_1 \otimes I_2)_A (I_1 \otimes (\sigma_x^r \sigma_z^s)_2)_B \, |\Psi\rangle_{A_i B_i} |\Psi\rangle_{A_j B_j}$$

Here, $\{\sigma_z, \sigma_x, \sigma_y\}$ are Pauli z, x and y operators, respectively. $(p,q)$ and $(r,s) \in \{0,1\}$ denote the Alice's and Bob's two-bit secret messages, respectively. Based on the secret message, the original state $|A_n B_n\rangle_i$ sent by Charlie is encoded into $|A_n B_n\rangle_i^e$ as shown in Table I and Table II by Alice and Bob, respectively.

Table 1: Change of states after encoding by Alice.

| Operations on $A_n A_m$ | Alice | Initial state sent by Charlie $|A_n B_n\rangle|A_m B_m\rangle$ | | | | |
|---|---|---|---|---|---|---|
| | Secret message | $|\phi^+\rangle$ | $|\phi^-\rangle$ | $|\psi^+\rangle$ | $|\psi^-\rangle$ | |
| $I \otimes I$ | 0000 | $|\phi^+\rangle|\phi^+\rangle$ | $|\phi^-\rangle|\phi^-\rangle$ | $|\psi^+\rangle|\psi^+\rangle$ | $|\psi^-\rangle|\psi^-\rangle$ | State after encoding $|A_n B_n\rangle_i^e |A_m B_m\rangle_j^e$ |
| $\sigma_z \otimes I$ | 0100 | $|\phi^-\rangle|\phi^+\rangle$ | $|\phi^+\rangle|\phi^-\rangle$ | $|\psi^-\rangle|\psi^+\rangle$ | $|\psi^+\rangle|\psi^-\rangle$ | |
| $\sigma_x \otimes I$ | 1000 | $|\psi^+\rangle|\phi^+\rangle$ | $-|\psi^-\rangle|\phi^-\rangle$ | $|\phi^+\rangle|\psi^+\rangle$ | $-|\phi^-\rangle|\psi^-\rangle$ | |
| $(\sigma_x \sigma_z = i\sigma_y) \otimes I$ | 1100 | $-|\psi^-\rangle|\phi^+\rangle$ | $|\psi^+\rangle|\phi^-\rangle$ | $-|\phi^-\rangle|\psi^+\rangle$ | $|\phi^+\rangle|\psi^-\rangle$ | |

Table II: Change of states after encoding by Alice by Bob.

| Operations on $B_n B_m$ | Bob | Initial state sent by Charlie $|A_n B_n\rangle|A_m B_m\rangle$ | | | | |
|---|---|---|---|---|---|---|
| | Secret message | $|\phi^+\rangle$ | $|\phi^-\rangle$ | $|\psi^+\rangle$ | $|\psi^-\rangle$ | |
| $I \otimes I$ | 00 | $|\phi^+\rangle|\phi^+\rangle$ | $|\phi^-\rangle|\phi^-\rangle$ | $|\psi^+\rangle|\psi^+\rangle$ | $|\psi^-\rangle|\psi^-\rangle$ | State after encoding $|A_n B_n\rangle_i^e |A_m B_m\rangle_j^e$ |
| $I \otimes \sigma_z$ | 01 | $|\phi^+\rangle|\phi^-\rangle$ | $|\phi^-\rangle|\phi^+\rangle$ | $-|\psi^+\rangle|\psi^-\rangle$ | $-|\psi^-\rangle|\psi^+\rangle$ | |
| $I \otimes \sigma_x$ | 10 | $|\phi^+\rangle|\psi^+\rangle$ | $|\phi^-\rangle|\psi^-\rangle$ | $|\psi^+\rangle|\phi^+\rangle$ | $|\psi^-\rangle|\phi^-\rangle$ | |
| $I \otimes i\sigma_y$ | 11 | $|\phi^+\rangle|\psi^-\rangle$ | $|\phi^-\rangle|\psi^+\rangle$ | $-|\psi^+\rangle|\phi^-\rangle$ | $-|\psi^-\rangle|\phi^+\rangle$ | |

**Step 4**: **Measurement**: Alice (Bob) measures her (his) qubits $A_i, A_j$ ($B_i, B_j$) in the Bell basis and obtain outcomes $a_{ij}$ ($b_{ij}$) respectively where $\{a_{ij}, b_{ij}\} \in \{00, 01, 10, 11\}$. They notify their measurement results to each other over an authenticated classical channel. Results of entanglement swapping for each two initial Bell-state are showed in Table 3 and Eq. 3. In this Table, entanglement swapping for each two arbitrary Bell-state are labeled to four groups $C_0$ to $C_3$ as [45]. As shown in Eq. (3), after entanglement swapping, each two initial Bell-state are converted to four unique combination that these combinations only are repeated once in each group $C_0$ to $C_3$.

$$
\begin{aligned}
C_0 &\to \{|\phi^+\rangle_{A_1A_2}|\phi^+\rangle_{B_1B_2}, |\phi^-\rangle_{A_1A_2}|\phi^-\rangle_{B_1B_2}, |\psi^+\rangle_{A_1A_2}|\psi^+\rangle_{B_1B_2}, |\psi^-\rangle_{A_1A_2}|\psi^-\rangle_{B_1B_2}\}, \\
C_1 &\to \{|\phi^-\rangle_{A_1A_2}|\phi^+\rangle_{B_1B_2}, |\phi^+\rangle_{A_1A_2}|\phi^-\rangle_{B_1B_2}, |\psi^+\rangle_{A_1A_2}|\psi^-\rangle_{B_1B_2}, |\psi^-\rangle_{A_1A_2}|\psi^+\rangle_{B_1B_2}\}, \\
C_2 &\to \{|\phi^+\rangle_{A_1A_2}|\psi^+\rangle_{B_1B_2}, |\phi^-\rangle_{A_1A_2}|\psi^-\rangle_{B_1B_2}, |\psi^+\rangle_{A_1A_2}|\phi^+\rangle_{B_1B_2}, |\psi^-\rangle_{A_1A_2}|\phi^-\rangle_{B_1B_2}\}, \\
C_3 &\to \{|\phi^-\rangle_{A_1A_2}|\psi^+\rangle_{B_1B_2}, |\phi^+\rangle_{A_1A_2}|\psi^-\rangle_{B_1B_2}, |\psi^-\rangle_{A_1A_2}|\phi^+\rangle_{B_1B_2}, |\psi^+\rangle_{A_1A_2}|\phi^-\rangle_{B_1B_2}\}.
\end{aligned}
\qquad (3)
$$

Table 3: outputs of entanglement swapping for each two arbitrary Bell-state.

| states | $|\phi^+\rangle_{A_2B_2}$ | $|\phi^-\rangle_{A_2B_2}$ | $|\psi^+\rangle_{A_2B_2}$ | $|\psi^-\rangle_{A_2B_2}$ |
|---|---|---|---|---|
| $|\phi^+\rangle_{A_1B_1}$ | $C_0$ | $C_1$ | $C_2$ | $C_3$ |
| $|\phi^-\rangle_{A_1B_1}$ | $C_1$ | $C_0$ | $C_3$ | $C_2$ |
| $|\psi^+\rangle_{A_1B_1}$ | $C_2$ | $C_3$ | $C_0$ | $C_1$ |
| $|\psi^-\rangle_{A_1B_1}$ | $C_3$ | $C_2$ | $C_1$ | $C_0$ |

**Step 5**: **Permission** If the controller wants the communication to proceed, he announces – over an authenticated channel– the two initial Bell states $\{A_iB_i, A_jB_j\}$ using four classical bits $\{c_{ii}, c_{jj}\} \in \{00, 01, 10, 11\}$ to Alice and Bob. Using this knowledge and the measurement results $b_{ij}(a_{ij})$ from the previous step, it becomes possible for Alice (Bob) to obtain $pq$ ($rs$) – the secret message from the other end. However, if the controller does not want the bidirectional direct communication to proceed, he simply does not reveal anything and none of the parties, have any information about the other parties' secret message.

**Example** Let us clarify the above protocol using an example. Let us assume that the two Bell states initially distributed by the controller are $|A_iB_i\rangle = |\phi^+\rangle_{A_iB_i}$ and $|A_jB_j\rangle = |\phi^-\rangle_{A_jB_j}$ and also let Alice's (Bob's) secret message be $pq = 00$ ($rs = 01$). After the encoding step the states become,

$$
|,\Omega\rangle = |\phi^+\rangle_{A_iB_i} \otimes |\phi^-\rangle_{A_jB_j} \xrightarrow[rs=01]{pq=00} |\phi^+\rangle_{A_iB_i} \otimes |\phi^+\rangle_{A_jB_j} = |\Omega\rangle^e
$$

$$
\begin{cases}
|\phi^+\rangle_{A_0A_1}|\phi^+\rangle_{B_0B_1} \\
|\phi^-\rangle_{A_0A_1}|\phi^-\rangle_{B_0B_1} \\
|\psi^+\rangle_{A_0A_1}|\psi^+\rangle_{B_0B_1} \\
|\psi^-\rangle_{A_0A_1}|\psi^-\rangle_{B_0B_1}
\end{cases} \to C_0
\qquad (4)
$$

After entanglement swapping and measuring of corresponding particles by Alice and Bob, the measurement results will be one of four terms in $C_0$ as shown in Eq. (4). For instance, if Alice's measurement result be $|\psi^+\rangle$ then Bob's measurement result will also be $|\psi^+\rangle$. Afterwards, they notify the measurement results to each other. After that, according to Table 3, they can understand that entanglement swapping is type $C_0$. At the permission step, Charlie notify the type of prepared initial states to Alice and Bob (in this case $|\phi^+\rangle_{A_0B_0} \otimes |\phi^-\rangle_{A_1B_1}$). Using this information, Alice and Bob know after encoding of their secret messages (00 and 01) $|A_iB_i\rangle = |\phi^+\rangle$ and $|A_jB_j\rangle = |\phi^+\rangle$ become $|A_iB_i\rangle^e = |\phi^+\rangle$ and $|A_jB_j\rangle^e = |\phi^-\rangle$, they can understand the main states sent by themselves and infer the secret messages that each of them has received from other. It is due to terms placed in each of the groups $C_0$ to $C_4$ are independent from other and the terms only are repeated once in each group.

## 4. Generalization

In the proposed method, larger entanglement states can be used for transmitting more information at the same time which is led to increasing of efficiency in the protocol. For this aim, GHZ state can be used instead of Bell states. So, the initial state can be written as the following:

$$\frac{1}{2}((|000\rangle + |111\rangle)(|000\rangle + |111\rangle))_{a0a1b0a2b1b2}$$

Which $a_0, a_1$ and $a_2$ belong to Alice and also, $b_0, b_1$ and $b_2$ belong to Bob.

In this case, they can encode their qubits as the following:

$$f(p,q,r,s,t,u,k,y)$$
$$\to ((\sigma_x^p\sigma_z^q)_1 \otimes (\sigma_x^r\sigma_z^s)_2 \otimes I_3)_A (I \otimes (\sigma_x^t\sigma_z^u)_2 \otimes (\sigma_x^k\sigma_z^y)_3)_B |\Psi\rangle_{A0_iA1_iB0_i}|\Psi\rangle_{A2_jB1_jB2_j}$$

Other steps are the same as explained in the previous one.

## 5. Proposed method with network

In this Section, a protocol of QSDC-based network is presented. This protocol communicates between three users in the network. It can be explained as the following scenario:

Suppose there are three users Alice, Bob and Elena with a supervisor _Charlie as the controller. Alice, Bob, and Elena want to communicate together and transmit their messages each to others so that each user only can receive and decode the corresponding message to himself/herself. Also, Charlie cannot understand the messages and only he can monitor the communication. This scenario can be described by the following protocol:

**Step 1**: **Distribution** The controller -Charlie- such as previous one, randomly prepares $2N + c$ GHZ states from the set of eight state $\{\phi\}_{ijk}$ $i, j, k \in \{0,1\}$ as shown in Eq. (5).

$$|\phi_{000}\rangle = \tfrac{1}{\sqrt{2}}(|000\rangle + |111\rangle),$$
$$|\phi_{001}\rangle = \tfrac{1}{\sqrt{2}}(|000\rangle - |111\rangle),$$
$$|\phi_{010}\rangle = \tfrac{1}{\sqrt{2}}(|100\rangle + |011\rangle),$$
$$|\phi_{011}\rangle = \tfrac{1}{\sqrt{2}}(|100\rangle - |011\rangle),$$
$$|\phi_{100}\rangle = \tfrac{1}{\sqrt{2}}(|010\rangle + |101\rangle), \quad (5)$$
$$|\phi_{101}\rangle = \tfrac{1}{\sqrt{2}}(|010\rangle - |101\rangle),$$
$$|\phi_{110}\rangle = \tfrac{1}{\sqrt{2}}(|110\rangle + |001\rangle),$$
$$|\phi_{111}\rangle = \tfrac{1}{\sqrt{2}}(|110\rangle - |001\rangle).$$

Where $N$ is the number of qubits used for encoding the message, $c$ is the numbers of qubits used for the security checking. Then, Charlie distributes the qubits $A_x$ to Alice, $B_x$ to Bob and $C_x$ to Elena ($x = 1,2,\ldots,2N + c$) and then each of them acknowledge the receipt of their respective qubits.

**Step 2**: **Channel verification** After Alice, Bob and Elena confirm that they received sequences, Alice executes security checking step as that she randomly selects $c$ 'checking' particles and measure them randomly in either $X$ or $Z$ basis. After that, she publishes the positions, measurement bases and measurement results to Bob and Elena by classical channels. Bob and Elena perform the same measurements in the same bases with Alice in corresponding particles. Alice also sends positions of this particles to Charlie. Then, Charlie notify the initial states of these 'checking' pairs to Alice, Bob and Charlie. Afterwards, they compare their measurement results. If there is no eavesdropper, measurement results related to users must be correlated. If the error rate goes beyond the threshold, they will abort the communication, else they will continue to the next step.

**Step 3**: **Encoding**: After verifying the security of quantum channel, from the remaining $2N$ pairs, Alice, Bob and Elena randomly and according to a previous agreement together, divide particles for transmitting their secret messages to each other. For example, Alice, Bob and Elena may simply select pairs divisible by three, respectively. Then, they randomly create three-particle groups $((A_i, B_i, C_i), (A_j, B_j, C_j))$ in their sequences. For example, Alice (Bob) or Elena may group particles $\{P_1(A), P_2(A)\}$ ($\{(P_1(B),P_2(B)\}$) or $\{P_1(C), P_2(C)\}$ in Group 1, particles $\{P_3(A),P_4(A)\}$ ($\{P_3(B),P_4(B)\}$) or $\{P_3(C),P_4(C)\}$ in Group 2 and so on. Then, in order to encode their two-bit secret message on the i-th group, Alice, Bob and Elena performs the following operation on their corresponding group,

$$f(p,q,r,s,k,y) \to ((\sigma_x^p \sigma_z^q)_1 \otimes I_2)_A (I_1 \otimes (\sigma_x^r \sigma_z^s)_2)_B (I_1 \otimes (\sigma_x^k \sigma_z^y)_2)_C |\Psi\rangle_{A_i B_i C_i} |\Psi\rangle_{A_j B_j C_j}$$

Or

$$f(p,q,r,s,k,y) \to ((\sigma_x^p \sigma_z^q)_1 \otimes I_2)_A ((\sigma_x^r \sigma_z^s)_1 \otimes I_2)_B ((\sigma_x^k \sigma_z^y)_1 \otimes I_2)_C |\Psi\rangle_{A_i B_i C_i} |\Psi\rangle_{A_j B_j C_j}$$

Other steps of the protocol are the same as the main protocol in the first section.

## 6. Security

In the proposed protocol, security analysis is only needed in the step of distributing the entangled pairs from Charlie to Alice-Bob. Due to using of entanglement swapping, the secret bits are not transmitted through the channel. So, it is only possible that Eve (eavesdropper) attack in step of preparation of channel (Step 1). In the rest of this section, three important types of security analyses are performed: controller permission review, review of control of eavesdropper and review of possible Attacks of an eavesdropper.

### 6.1. Security analysis without the controller's permission

Lack of receiving information by the receiver without the controller's permission is one of the important issues in Bidirectional Controlled Quantum Secure Direct Communication (BCQSDC) protocols. Sending information in the proposed protocol is bidirectional, i.e., two users can send information to each other, simultaneously. But for obtaining information and decoding them, users need to know the information about the initial states sent by the controller. Without receiving this information, they cannot get any part of the secret messages. For example, consider the first example stated in this paper where Alice and Bob send secret messages 00 and 01 to each other simultaneously. For detecting their secret messages if they do not know initial state, according to Tables 1, 2 and 3, because of the distribution of same probability between initial states they cannot detect secret messages as shown in Eq. (5).

$$P(xy = 00|kl) = P(xy = 01|kl) = P(xy = 10|kl) = P(xy = 11|kl) = \frac{1}{16} \quad (5)$$
$$kl \in \{(00), (01), (10), (11)\}$$

Here, $P(xy = 11|kl)$ is the probability that the initial bell state is xy = 11 that given the secret message was $kl$. Security analysis with showing all of the classical information.

Also, if Eve achieves all of the classical information transmitted by users and controller. She cannot get any part of the secret messages. Because she needs to mental knowledge of users to their secret messages i.e., Alice and Bob only themselves know about their secret messages that had sent to each other. For example, in the first example, Alice and Bob know secret messages sent by themselves are 00 and 01, respectively. And they only know this information and no anyone. We can present this case with the same probabilities as shown in Eq. (6).

$$prob(kl = 00|xymnop) = prob(kl = 01|xymnop) = prob(kl = 10|xymnop) = prob(kl = 11|xymnop) = \frac{1}{4}$$
$$kl, mn, op \in \{(00), (01), (10), (11)\}$$
(6)

Where variable $Prob$ indicate the probability of achieving to the secret message sent by Alice (Bob). Also, $kl$ indicate the secret message sent by Alice (Bob) and also $xy$ and $mn$ present classical information sent by Alice and Bob to each other. In addition, $op$ show classical information sent by Charlie to Alice and Bob.

### 6.2. Security analysis against Eve's Attacks

This section shows that the proposed protocol is secure against three types of attacks: 1) intercept-and-resend attack 2) controlled- not attack, and 3) entanglement-and-measure attack [46, 47]. These attacks are described with details in the following:

### 6.2.1. Intercept-and-resend attack

In this attack, Eve may attempt to play the role of Alice or Bob to read the secret message. In this case, the presence of eavesdropper can be realized in security checking (the second step) since there are no correlations at all between the qubits Alice (Bob), Charlie's classical information and the fake ones.

### 6.2.2. Controlled-not attack

In this attack, Eve wants to achieve secret messages by adding additional qubits to channel and measuring them. For this aim, she intercepts the qubits sent from Charlie to Alice (Bob) and then performs controlled-not gate to add one additional qubit on each EPR state. This attack can be detected in security checking step or second step. Because there are not correlations between particles again.

### 6.2.3. Entanglement-and-measure attack

In this attack, Eve wants to achieve partial information by entangling her auxiliary qubit, $|E_i\rangle$, with a qubit $a$ ($b$) before Alice (Bob) received it (assumed to be in the state $|i\rangle$, with $i \in \{0,1\}$). Assume Eve uses unitary operation $P_{NE}$, on the pairs $a(b)$ and her qubit, $E$. We can describe it as Eq. (7).

$$P_{NE}|i\rangle|E_i\rangle = \alpha|i\rangle|E_i\rangle + \beta|i\oplus1\rangle|E_{i\oplus1}\rangle, \quad |\alpha|^2 + |\beta|^2 = 1 \tag{7}$$

Where $\langle E_i|E_{i\oplus1}\rangle = 0$ and $N$ is Alice's (Bob's) qubit. Then Eve resends $N$ to Alice (Bob). In this case, Eve will be realized with a probability $|\beta|^2$ in security checking process in the second step.

## 7. Comparison

In this paper, we present a Bidirectional Controlled Quantum Secure Direct Communication (BCQSDC) protocol using EPR pairs via entanglement swapping for transmitting secret messages. Unlike previous works [34, 38, 41], in this work, two users can transmit two-qubit to each other, simultaneously. In this work, we use lower resources into previous works.

For analysis and comparison of efficiency QSDC protocols, the bellow definition is used [41, 48].

$$\eta_1 = \frac{m_u}{q_k + b_k}, \tag{8}$$

Where $m_u$, $q_k$ and $b_k$ indicate the number of secret bits, the number of total qubits and the number of classical bits respectively. Since transmission of qubits is more complex and expensive than the classical bits, we can redefine the efficiency as Eq. (9).

$$\eta_2 = \frac{m_u}{q_k} \tag{9}$$

Table 4 shows a comparison of the efficiency of the proposed protocol with the previous schemes. It is obvious that the efficiency of the proposed protocol is higher than previous protocols. Also, previous protocols are one way and they for converting their protocols to bidirectional, need to run their protocol twice [32, 41], i.e., they need to the number of qubits same with Alice for Bob to create a protocol for sending qubits from Bob to Alice named so-called a quantum dialogue (QD). But our protocol is bidirectional.

Furthermore, the number of security checking phases in our protocol and in [6] is one. Also, in our protocol and in [6] qubits aren't send directly and these protocols used from the feature of entanglement swapping into other works. So, the number of security checking steps is reduced. But, our work is very better than [6] in term of efficiency and our method is bidirectional. But the method in [6] is one way. Bidirectional means that Alice and Bob can transmit a two-qubit state to each other, simultaneously under permission of a controller by one entangled state as a quantum channel.

| Protocol | Type | $m_u$ | $q_k$ | $b_k$ | $\eta_1$ (%) | $\eta_2$ (%) | The number of Security Check | The number of channels | Direct sending of qubits |
|---|---|---|---|---|---|---|---|---|---|
| [32] | BCQSDC | 4 | 8 | 4 | 33.33 | 50 | 3 | 2 | Yes |
| [34] | CQSDC | 2 | 6 | 3 | 22.22 | 33.33 | 1 | One way | No |
| [41] | BQSDC | 4 | 8 | 4 | 33.33 | 50 | 2 | 2 | Yes |
| Our Method | BCQSDC | 4 | 4 | 8 | 33.33 | 100 | 1 | 1 | No |

**Conclusion**

In this paper, a Controlled Bidirectional Quantum Secure Direct Communication (CBQSDC) was presented that users can transmit two-bit message using EPR pairs to each other simultaneously under the permission of the controller. Also, in this method entanglement swapping was used for transmitting the secret message. So, no information is directly transmitted in the quantum channel. Due to using of entanglement swapping, the number of security checking steps was reduced. Also, in the proposed protocol, the number of used quantum resources was reduced into previous ones. The proposed protocol effectively improved the efficiency of the previous ones.